# Leveraging Sociological Models for Predictive Analytics

Richard Colbaugh
Sandia National Laboratories
Albuquerque, NM USA
colbaugh@comcast.net

Kristin Glass
New Mexico Tech
Socorro, NM USA
kglass@icasa.nmt.edu

Travis Bauer
Sandia National Laboratories
Albuquerque, NM USA
tlbauer@sandia.gov

*Abstract*—There is considerable interest in developing techniques for predicting human behavior, for instance to enable emerging contentious situations to be forecast or the nature of ongoing but "hidden" activities to be inferred. A promising approach to this problem is to identify and collect appropriate empirical data and then apply machine learning methods to these data to generate the predictions. This paper shows the performance of such learning algorithms often can be improved substantially by leveraging sociological models in their development and implementation. In particular, we demonstrate that sociologically-grounded learning algorithms outperform gold-standard methods in three important and challenging tasks: 1.) inferring the (unobserved) nature of relationships in adversarial social networks, 2.) predicting whether nascent social diffusion events will "go viral", and 3.) anticipating and defending future actions of opponents in adversarial settings. Significantly, the new algorithms perform well even when there is limited data available for their training and execution.

## I. Introduction

There is great interest in developing techniques for accurately predicting human behavior. For example, forecasting the eventual outcomes of social processes is a central concern in domains ranging from popular culture to public policy to national security [1]. The task of inferring the existence and nature of activities which are presently underway but not directly observable, sometimes referred to as "predicting the present" [2], is also of crucial importance in many applications. A promising approach to obtaining such predictions is to identify and collect empirical data which appropriately characterize the phenomenon of interest and then to analyze these data using machine learning (ML) methods [3]. Roughly speaking, ML algorithms automatically "learn" relationships between observed variables from examples presented in the form of training data; the learned relationships are then used to generate predictions in new situations. ML's capacity to learn from examples, scale to large datasets, and adapt to new or changing conditions make this an attractive approach to predictive analysis.

The work reported in [4-13] illustrates some of the ways ML can be used for forecasting, and in particular how these techniques can be applied to online (Web) data in order to predict the outcomes of a broad range of social dynamics processes (e.g., social movements, cultural and financial markets, protest events, political elections, and economic activity). Alternatively, the papers [14-20] derive ML techniques for predicting the present, for instance enabling the existence of hidden links in social networks to be inferred, the sentiment of informal communications to be estimated, and the spread of various health-related phenomena to be remotely monitored.

Existing ML methods, although very useful, face at least two challenges. First, the prediction accuracy obtainable even with state-of-the-art algorithms is sometimes insufficient for the task at hand, such as when the predictions are to be used to inform high-consequence decisions (e.g., pertaining to national security or human health). Second, applying ML techniques typically requires that significant quantities of data be collected and "labeled". For example, deriving an ML scheme for estimating the sentiment polarity of blog posts usually involves collecting, processing, and manually labeling hundreds of example posts expressing positive and negative sentiment [15]. Employing ML for forecasting ordinarily entails assembling extensive time series traces, implying that such methods may not be responsive enough to generate useful predictions about rapidly emerging events [13]. Additionally, realizing good performance with standard ML usually necessitates frequent retraining to permit algorithms to adapt to evolving conditions, which limits usefulness in many domains (e.g., in adversarial settings in which opponents adapt their behaviors expressly to defeat learning algorithms [21]).

This paper proposes that the challenges of predicting human behavior using ML often can be overcome by leveraging sociological models in the development and implementation of the learning algorithms. This proposal is motivated by our recent research showing that including sociologically-meaningful measures of network dynamics as features in ML algorithms permits predictions regarding social dynamics which are substantially more accurate than those based on standard features [22]. The present paper initiates a more systematic exploration of the utility of combining ML with sociological models for social prediction. In particular, we consider three important and challenging tasks: 1.) inferring the "signs" (i.e., friendly or antagonistic) of relationship ties in social networks, 2.) predicting whether a nascent social diffusion event of interest will ultimately propagate widely or will instead quickly dissipate, and 3.) anticipating and countering the future actions of opponents in adversarial settings.

The paper makes four primary contributions. First, we consider the problem of predicting edge-signs in social networks, where positive/negative edges reflect friendly/antagonistic social ties, and derive a novel ML algorithm for edge-sign prediction which leverages structural balance theory [23-25]. The proposed algorithm outperforms a "gold-standard" method in empirical tests with two large-scale online social networks, with the boost in prediction accuracy being especially significant in situations where only limited training data are available. Interestingly, the inferred edge-signs are also shown to be useful when predicting the way adversarial networks will fracture under stress. Second, we examine the problem of forecasting the ultimate reach of "complex contagion" events [26,27]. A predictability assessment of complex contagion dynamics indicates that metrics which should be predictive of a contagion's reach are fairly subtle measures of the network dynamics associated with very early diffusion activity. These results are used to derive an ML algorithm for predicting which complex contagion events will ultimately "go viral" and which won't, and it is demonstrated that the algorithm outperforms standard methods in an empirical investigation of online meme propagation [28]. Significantly, the new algorithm performs well even when only limited time series data are available for analysis, permitting reliable predictions early in the contagion lifecycle. It is also shown that the proposed algorithm enables effective early warning analysis for an important class of cyber threats.

Third, we consider the problem of predicting the actions of opponents in adversarial settings [21], and combine ML with a game-theoretic model [29,30] to a design predictive defense system which is capable of countering current and future adversary behaviors. An empirical study employing an extensive set of cyber attack data shows that the proposed predictive defense is much more effective than existing defense systems. For instance, compared with standard cyber defenses, the predictive method is better able to distinguish malicious attacks from legitimate behaviors, allows successful defense against completely new attacks, and requires significantly less retraining to maintain good performance in evolving threat conditions.

Finally, and more generally, the set of results presented in this paper suggests that incorporating even simple concepts and models from sociology can substantially improve the performance of ML prediction algorithms. These "sociology-aware" algorithms outperform gold-standard methods, particularly when there is limited data available for training and implementing the methods, and can be applied at no additional cost. Thus this initial investigation provides compelling support for the proposal that sociological models can be usefully leveraged in the design and implementation of predictive analytics methods.

We now present our results on inferring edge-signs (Section II), forecasting contagion events (Section III), and designing proactive defenses (Section IV). For each application we begin by stating the prediction problem of interest and briefly reviewing the sociological model we use in its solution. We then derive an ML prediction algorithm which is based, in part, on this sociological model, and evaluate the performance of the proposed algorithm through an empirical case study. Additionally, it is shown how these analysis algorithms can be applied to key security informatics challenges.

## II. PREDICTING LINK-SIGNS

### A. Problem Formulation

Social networks may contain both positive and negative relationships – people form ties of friendship and support but also of animosity or disapproval. These two types of social ties can be modeled by placing signs on the links or edges of the social network, with +1 and −1 reflecting friendly and antagonistic relationships, respectively. We wish to study the problem of predicting the signs of certain edges of interest by observing the signs and connectivity patterns of the neighboring edges. More specifically, for a directed social network $G_s = (V, E)$ with signed edges, where V and E are the vertex and edge sets, respectively, we consider the following edge-sign prediction problem: given an edge $(u,v) \in E$ of interest for which the edge-sign is "hidden", infer the sign of (u,v) using information contained in the remainder of the network.

It is natural to suspect that *structural balance theory* (SBT) may be useful for edge-sign prediction. Briefly, SBT posits that if $w \in V$ forms a triad with edge (u,v), then the sign of (u,v) should be such that the resulting signed triad possessing an odd number of positive edges; this encodes the common principle that "the friend of my friend is my friend", "the friend of my enemy is my enemy", and so on [23,24]. Thus SBT suggests that knowledge of the signs of the edges connecting (u,v) to its neighbors may be useful in predicting the sign of (u,v).

### B. Prediction Algorithm

We approach the task of predicting the sign of a given edge (u,v) in the social network $G_s$ as an ML classification problem. The first step is to define, for a given edge, a collection of features which may be predictive of the sign of that edge. To allow a comparison with the (gold-standard) prediction method given in [25], we adopt the same two sets of features used in that study. For a given edge (u,v), the first set of features defined in [25] characterize the various triads to which (u,v) belongs. Because triads are directed and signed, there are sixteen distinct types (e.g., the triad composed of positive edge (u,w) and negative edge (w,v), together with (u,v), is one type). Thus the first sixteen features for edge (u,v) are the counts of each of the various triad types to which (u,v) belongs. Including these features is directly motivated by SBT. For example, if (u,v) belongs to many triads with one positive and one negative edge, it may be likely that the sign of (u,v) is negative, since then these triads would possess an odd number of positive edges and therefore be "balanced".

The second set of features defined in [25] measure characteristics of the degrees of the endpoint vertices u and v of the given edge (u,v). There are five of these features, quantifying the positive and negative out-degrees of u, the positive and negative in-degreed of v, and the total number of neighbors u and v have in common (interpreted in an undirected sense). Combining these five measures with the sixteen triad-related features results in a feature vector $x \in \Re^{21}$ for each edge of interest (see [25] for a more thorough discussion of these features and the motivation for selecting them). The feature vector x associated with an edge (u,v) will form the basis for predicting the sign of that edge.

We wish to learn a vector $c \in \Re^{21}$ such that the classifier orient = $\text{sign}(c^T x)$ accurately estimates the sign of the edge whose features are encoded in vector x. Vector c is learned, in part, from labeled examples of positive and negative edges. Additionally, the proposed learning algorithm leverages the insights of SBT. A simple way to incorporate SBT is to assemble sets $V^+$ and $V^-$ of positive and negative features, that is, sets of features which according to SBT ought to be associated with positive and negative edges, respectively. The triads to which (u,v) belongs in which the other two edges are positive are predicted by SBT to "contribute" to (u,v) being positive; thus the four features corresponding to triads with two positive labeled edges are candidates for membership in $V^+$ (there are four such features because $G_s$ is directed). Analogously, SBT posits that the eight features indexing triads in which exactly one of the two edges that neighbor (u,v) is positive are candidates for membership in $V^-$. (Note that the remaining four triad features index triads in which both of the edges neighboring (u,v) are negative, and as there is less empirical support for SBT in this case [25] these features are not assigned to either $V^+$ or $V^-$.)

We now derive an ML algorithm for edge-sign prediction which is capable of leveraging SBT in its learning process. The development begins by modeling the problem data as a bipartite graph $G_b$ of edge-sign instances and features (see Figure 1). If there are n edges and 21 features, it can be seen that the adjacency matrix A for graph $G_b$ is given by

$$A = \begin{bmatrix} 0 & X \\ X^T & 0 \end{bmatrix} \quad (1)$$

where matrix $X \in \Re^{n \times 21}$ is constructed by stacking the feature vectors $x_i$ as rows, and each '0' is a matrix of zeros.

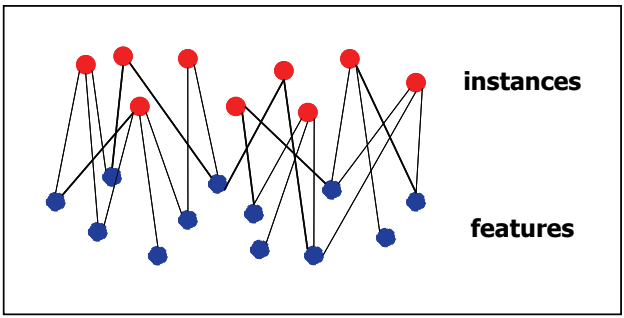

**Figure 1.** Cartoon of bipartite graph data model $G_b$, in which edge-instances (red vertices) are connected to the features (blue vertices) they contain, and link weights (black edges) reflect the magnitudes taken by the features in the associated instances.

Assume the initial problem data consists of a set of n edges, of which $n_l \leq n$ are labeled, and a set of labeled features $V_l = V^+ \cup V^-$, and suppose this label information is encoded as vectors $d \in \Re^{n_l}$ and $w \in \Re^{|V_l|}$, respectively. Let $d_{est} \in \Re^n$ be the vector of estimated signs for the edges in the dataset, and define the "augmented" classifier $c_{aug} = [d_{est}^T \; c^T]^T \in \Re^{n+21}$ that estimates the polarity of both edges and features. Note that the quantity $c_{aug}$ is introduced for notational convenience and is not directly employed for classification. More specifically, in the proposed methodology we learn $c_{aug}$, and therefore c, by solving an optimization problem involving the labeled and unlabeled training data, and then use c to estimate the sign of any new edge of interest with the simple classifier orient=$\text{sign}(c^T x)$. Assume for ease of notation that the edges and features are indexed so that the first $n_l$ elements of $d_{est}$ and $|V_l|$ elements of c correspond to the labeled data.

We wish to learn an augmented classifier $c_{aug}$ with the following three properties: 1.) if an edge is labeled, then the corresponding entry of $d_{est}$ should be close to this ±1 label; 2.) if a feature is in the set $V_l = V^+ \cup V^-$, then the corresponding entry of c should be close to this ±1 polarity; and 3.) if there is an edge $X_{ij}$ of $G_b$ that connects an edge x and a feature f and $X_{ij}$ possesses significant weight, then the estimated polarities of x and f should be similar. These objectives are encoded in the following optimization problem:

$$\min_{c_{aug}} c_{aug}^T L c_{aug} + \beta_1 \sum_{i=1}^{n_l} (d_{est,i} - d_i)^2 + \beta_2 \sum_{i=1}^{|V_l|} (c_i - w_i)^2 \quad (2)$$

where $L = D - A$ is the graph Laplacian matrix for $G_b$, with D the diagonal degree matrix for A (i.e., $D_{ii} = \Sigma_j A_{ij}$), and $\beta_1$, $\beta_2$ are nonnegative constants. Minimizing (2) enforces the three properties we seek for $c_{aug}$, with the second and third terms penalizing "errors" in the first two properties. To see that the first term enforces the third property, observe that this expression is a sum of components of the form $X_{ij}(d_{est,i} - c_j)^2$. The constants $\beta_1$, $\beta_2$ are used to balance the relative importance of the three properties. The $c_{aug}$ which minimizes objective function (2) can be obtained by solving the following set of linear equations:

$$\begin{bmatrix} L_{11} + \beta_1 I_{nl} & L_{12} & L_{13} & L_{14} \\ L_{21} & L_{22} & L_{23} & L_{24} \\ L_{31} & L_{32} & L_{33} + \beta_2 I_{|V_l|} & L_{34} \\ L_{41} & L_{42} & L_{43} & L_{44} \end{bmatrix} c_{aug} = \begin{bmatrix} \beta_1 d \\ 0 \\ \beta_2 w \\ 0 \end{bmatrix} \quad (3)$$

where the $L_{ij}$ are matrix blocks of L of appropriate dimension.

We summarize this discussion by sketching an algorithm for learning the proposed edge-sign prediction (ESP) classifier:

**Algorithm ESP**

1. Construct the set of equations (3).
2. Solve equations (3) for $c_{aug} = [\; d_{est}^T \; c^T \;]^T$ (for instance using the Conjugate Gradient method).
3. Estimate the sign of any new edge x of interest as: orient = $\text{sign}(c^T x)$.

The utility of Algorithm ESP is now examined through a case study involving edge-sign estimation for two social networks extracted from the Wikipedia online encyclopedia.

## C. Wikipedia Case Study

This case study examines the performance of Algorithm ESP for the problem of estimating the signs of the edges in two social networks extracted from Wikipedia (WP), a collectively-authored online encyclopedia with an active user community. We consider the following WP social networks: 1.) the graph of 103,747 edges corresponding to votes cast by WP users in elections for promoting individuals to the role of 'admin' [25], and 2.) the graph of 740,397 edges characterizing editor interactions in WP [31]. In each network, the majority of the edges ($\approx$80%) are positive. Thus we follow [25] and create balanced datasets consisting of 20K positive and 20K negative edges for the "voting" network [25], and 50K positive and 50K negative edges for the "interaction" network [31].

This study compares the edge-sign prediction accuracy of Algorithm ESP with that of the impressive gold-standard logistic regression classifier given in [25]. The gold-standard algorithm is applied exactly as described in [25]. Algorithm ESP is implemented with parameter values $\beta_1 = 0.1$ and $\beta_2 = 0.5$, and with the vector w constructed using the four "positive triad" features $V^+$ and eight "negative triad" features $V^-$ noted above. As a focus of the investigation is evaluating the extent to which good prediction performance can be achieved even when only a limited number of labeled edges are available for training, we examine training sets which incorporate a range of numbers of labeled edges: $n_l$ = 0, 10, 20, 50, 100, 200.

Sample results from this study are depicted in Figures 2 and 3. Each data point in the plots represents the average of ten trials. In each trial, the edges are randomly split into equal-size training and testing sets, and a randomly selected subset of the training edges of size $n_l$ is "labeled" (i.e., the labels for these edges are made available to the learning algorithms). It can be seen that Algorithm ESP outperforms the gold-standard method on both datasets, and that the improved accuracy obtained with the proposed "SBT-informed" algorithm is particularly significantly when the number of labeled training instances is small.

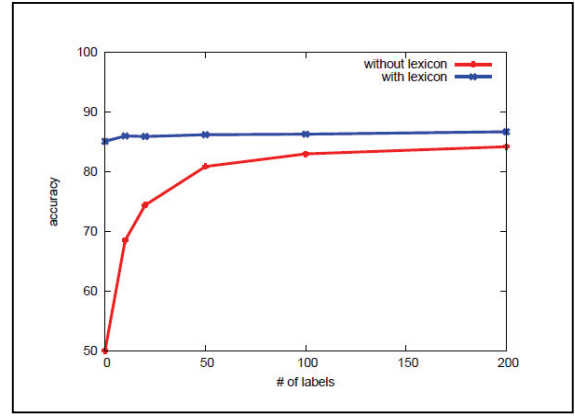

**Figure 3.** Results for WP "interaction network" case study. The plot shows how edge-sign prediction accuracy (vertical axis) varies with the number of available labeled training instances (horizontal axis) for two classifiers: gold-standard (red) and Algorithm ESP (blue).

## D. Network Fracture Case Study

Recently it has been proposed that structural balance theory can be used to predict the way a network of entities (e.g., individuals, countries) will split if subjected to stress [32], a capability of relevance in many security applications. Briefly, [32] models the polarity and intensity of relationships between the entities of interest as a completely connected network with weighted adjacency matrix $Z=Z^T \in \Re^{n \times n}$, where matrix element $z_{ij}$ represents the strength of the friendliness or unfriendliness between entities i and j. Note that this network model is somewhat more general than the one introduced above, in that each edge relating two individuals possesses both a sign and an intensity.

SBT is a "static" theory, positing what a stable configuration of edge-signs in a social network should look like. However, underlying the theory is a dynamical idea of how unbalanced network triads ought to resolve themselves to become balanced. A model which captures this underlying dynamics is given by the simple matrix differential equation [32]

$$dZ/dt = Z^2, \quad Z(0)=Z_0. \quad (4)$$

To see the connection between these dynamics and SBT, observe that (4) specifies the following dynamics for entry $z_{ij}$:

$$dz_{ij}/dt = \Sigma_k z_{ik} z_{kj}.$$

Thus if triad {i,j,k} is such that $z_{ik}$ and $z_{kj}$ have the same sign, the participation of $z_{ij}$ in this triad will drive $z_{ij}$ in the positive direction, while if they have opposite signs then $z_{ij}$ will be driven in the negative direction. These dynamics therefore favor triads with an odd number of positive edge-signs, consistent with SBT [23].

The paper [32] proves that, for generic initial conditions $Z_0$, system (4) evolves to a balanced pattern of edge-signs in finite time; the balanced configuration is guaranteed to be composed of either all positive edges or two all-positive cliques connected entirely by negative edges. These configurations can be interpreted as predictions of the way a social network described by $Z_0$ will fracture if subjected to sufficient stress. More precisely, given a model $Z_0$ for a signed social network, model (4) can be used as the basis for the following two-step procedure for predicting the way the network will fracture: 1.) integrate (4) forward in time until it reaches singularity $Z_s$ (this singularity will be reached in finite time), and 2.) interpret $Z_s$ as defining a split of the network into two groups, where each group has all positive intra-group edges and the inter-group edges are all negative (and where one of the groups could be empty). See Figure 4 for an illustration of the dynamics of system (4).

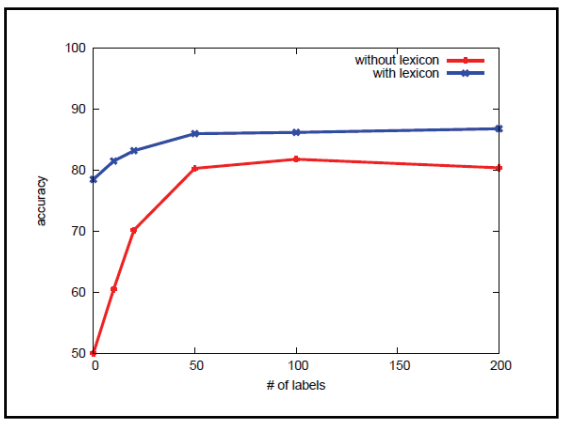

**Figure 2.** Results for WP "voting network" case study. The plot shows how edge-sign prediction accuracy (vertical axis) varies with the number of available labeled training instances (horizontal axis) for two classifiers: gold-standard (red) and Algorithm ESP (blue).

Remarkably, [32] shows that predictions obtained in this manner are in excellent agreement with two real-world cases of group fracture for which there is empirical data: the division of countries into Allied and Axis powers in World War II [33], and the split of the well-studied Zachary Karate Club into two smaller clubs [34]. However, the analysis presented in [32] requires that matrix $Z_0$ be completely known, that is, that all of the "initial" relationships $z_{ij}(0)$ between entities be measurable. Such comprehensive data are not always available in practical applications.

We have found that the requirement that relationship matrix $Z_0$ be perfectly known can be relaxed through the use of Algorithm ESP. More specifically, given a subset of the relationship data, the remaining weighted edge-signs can be estimated using Algorithm ESP, and these estimates $\underline{Z}_0$ can be used in place of $Z_0$ when initializing (4). We have tested this procedure using the relationship network proposed in [33] for 17 key countries involved in World War II. This investigation demonstrates that accurate prediction of which countries would eventually join the Allied forces and which would become Axis members can be made with less than 15% of the edge-signs known in advance. For example, data for only the relationships maintained by Germany and the USSR is sufficient to enable correct prediction of the ultimate alignment of all countries except Portugal (see Figure 4). Similar results hold for analysis of the split of the Zachary Karate Club [34].

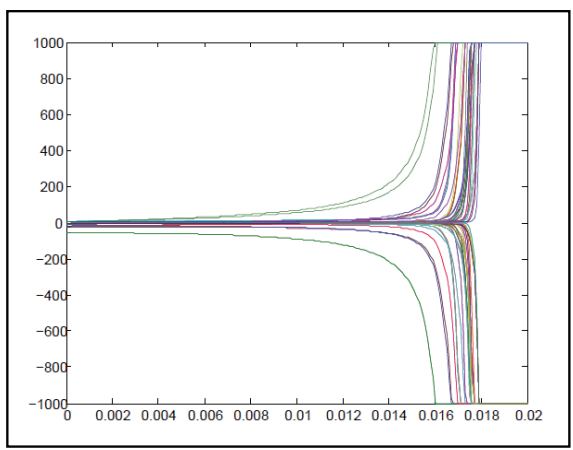

**Figure 4.** SBT dynamics. The evolution of model (4) initialized at the (scaled) "propensity" matrix given in [33] (horizontal axis is time and vertical axis is edge-weight).

## III. EARLY WARNING FOR COMPLEX CONTAGIONS

### A. Problem Formulation

There is significant interest in developing predictive capabilities for social diffusion processes, for instance to permit early identification of emerging contentious situations or accurate forecasting of the eventual reach of potentially "viral" behaviors. This section considers the diffusion early warning problem: we suppose some sort of triggering event has taken place and wish to determine, as early as possible, whether this event will ultimately generate a large, self-sustaining reaction, involving the propagation of behavioral changes through a substantial portion of a population, or will instead quickly dissipate. Of particular interest is the propagation of behaviors that are costly or controversial, or about which there is uncertainty, as the diffusion of such activities often have significant societal impact [13].

Recent research has shown that such behaviors may spread as *complex contagions*, requiring social affirmation or reinforcement from multiple sources in order to propagate [26,27]. Because the diffusion dynamics for complex contagions are different than those of "simple" contagions like disease epidemics, it is natural to suspect that developing effective early warning algorithms for complex contagions may require careful consideration of these more complex dynamics. In this section we explore this possibility by deriving an early warning method for complex contagions which explicitly leverages a mathematical model for these diffusion events. We adopt the contagion model proposed in [27], implemented on a class of social networks which possess realistic topologies, and analyze this model to identify features of the contagion that are likely to be predictive of diffusion reach. These features are then used as the basis for an ML algorithm which distinguishes complex contagions that will propagate widely from those which will quickly dissipate.

### B. Predictability Assessment

Here we briefly describe the results of applying the predictability assessment procedure presented in [1,13] to the task of identifying measurables that should be predictive of complex contagion success. The discussion begins with short, intuitive reviews of our predictability assessment process and network diffusion modeling framework, and then summarizes the main results obtained via this theoretical analysis.

**Predictability.** The basic idea behind the proposed approach to predictability analysis is simple and natural: we assess predictability by answering questions about the reachability of diffusion events. To obtain a mathematical formulation of this strategy, the behavior about which predictions are to be made is used to define the system *state space subsets of interest* (SSI), while the particular set of candidate measurables under consideration allows identification of the *candidate starting set* (CSS), that is, the set of states and system parameter values which represent initializations that are consistent with, and equivalent under, the presumed observational capability. As a simple example, consider an online market with two products, A and B, and suppose the system state variables consist of the current market share for A, ms(A), and the rate of change of this market share, r(A) (ms(B) and r(B) are not independent state variables because ms(A) + ms(B) = 1 and r(A) + r(B) = 0); let the parameters be the advertising budgets for the products, b(A) and b(B). The producer of A might find it useful to define the SSI to reflect market share dominance by A, that is, the subset of the two-dimensional state space where ms(A) exceeds a specified threshold. If only market share and the advertising budgets can be measured then the CSS is the one-dimensional subset of state-parameter space consisting of the initial magnitudes for ms(A), b(A), and b(B), with r(A) unspecified.

Roughly speaking, the approach to predictability assessment proposed in [1,13] involves determining how probable it is to reach the SSI from a CSS and deciding if these reachability properties are compatible with the prediction goals. If a system's reachability characteristics are compatible with the prediction objectives the situation is deemed predictable (and otherwise it is unpredictable). This setup permits the identification of candidate predictive measurables: these are the measurable states and/or parameters which most strongly affect the predictability properties [1]. Continuing with the online market example, if trajectories with positive early market share rates r(A) are much more likely to yield market share dominance for A than are trajectories with negative early r(A), independent of the early values for ms(A), then the situation is unpredictable (because r(A) is not measured). Adding the capacity to measure r(A) would then increase system predictability, and depending upon the task requirements this new measurement ability could result in a predictable situation. A quantitative, mathematically-rigorous presentation of this predictability assessment framework can be found in [1,13].

**Model.** In complex contagion events, the probability of adopting a controversial or unproven behavior or idea increases with the number of other adopting *individuals*, and not merely the number of exposures to the contagion (so that multiple interactions with the same adopting individual do not increase the likelihood of adoption, as it does in simple contagions) [26,27]. Recently the authors of [27] proposed an empirically-grounded model for complex contagions in which individuals interact via a social network of arbitrary topology, and the probability that individual A adopts a given activity or idea is a function of the number of A's adopting neighbors; the functional form of this adoption "influence curve" is obtained empirically (see [27] for a detailed description of the model).

The dynamics of contagion may depend upon the topological structure of the underlying social network. This dependence suggests that, in order to identify the features of complex contagions which have predictive power, it is necessary to assess predictability using social network models with realistic topologies. Therefore in this study we implement the complex contagion model [27] with social networks that possess four topological properties which are ubiquitous in the real-world [13]: right-skewed degree distribution, transitivity, community structure, and core-periphery structure.

It is shown in [1] that *stochastic hybrid dynamical systems* (S-HDS) provide a useful mathematical formalism with which to represent social contagions on realistic networks (see Figure 5). An S-HDS is a feedback interconnection of a discrete-state stochastic process, such as a Markov chain, with a family of continuous-state stochastic dynamical systems [1]. Combining discrete and continuous dynamics within a unified, computationally tractable framework offers an expressive, scalable modeling environment that is amenable to formal mathematical analysis. In particular, S-HDS models can be used to efficiently represent and analyze social contagion on large-scale networks with the four topological properties listed above [13].

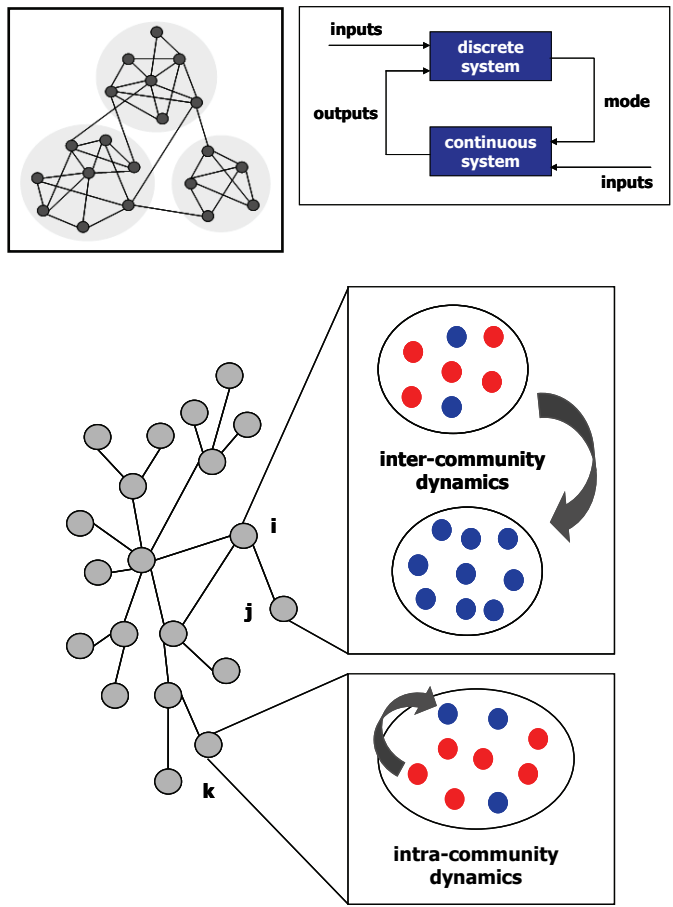

**Figure 5.** Modeling complex contagion on networks with community structure via S-HDS. The cartoon at top left depicts a network with three communities. The cartoon at bottom illustrates contagion *within* a community k and *between* communities i and j. The schematic at top right shows the basic S-HDS feedback structure.

As an intuitive illustration of the way S-HDS enable effective, tractable representation of complex contagion phenomena, consider the task of modeling contagion on a network possessing community structure. As shown in Figure 5, the contagion proceeds in two ways: 1.) *intra-community diffusion*, involving frequent interactions between individuals within the same community and the resulting gradual change in the concentrations of adopting (red) individuals, and 2.) *inter-community diffusion*, in which the "infection" jumps from one community to another, for instance because an adopting individual encounters a new community. S-HDS models offer a natural framework for representing these dynamics, with the S-

HDS continuous system modeling the intra-community dynamics (e.g., via stochastic differential equations), the discrete system capturing inter-community dynamics (e.g., using a Markov chain), and the interplay between these dynamics being encoded in the S-HDS feedback structure (e.g., the transition probabilities of the discrete system Markov chain may depend upon the state of the continuous system) [13].

**Results.** We applied the predictability assessment methodology summarized above to a "realistic network" version of the complex contagion model given in [27] (i.e., the model obtained by implementing the dynamics specified in [27] on a class of networks possessing the four topological properties summarized above). The main finding of this study is that the predictability of the reach of complex contagions depends crucially upon the social network's community and core-periphery structures. These findings are now summarized more quantitatively.

We adopt a modularity-based definition for network community structure [35], whereby a good partitioning of a network's vertices into communities is one for which the number of edges between putative communities is smaller than would be expected in a random partitioning. To be concrete, a modularity-based partitioning of a network into two communities maximizes the modularity $Q = s^T B s / 4m$, where m is the total number of edges in the network, the partition is specified with the elements of vector s by setting $s_i = 1$ if vertex i belongs to community 1 and $s_i = -1$ if it belongs to community 2, and the matrix B has elements $B_{ij} = A_{ij} - k_i k_j / 2m$, with $A_{ij}$ and $k_i$ denoting the network adjacency matrix and degree of vertex i, respectively. Partitions of the network into more than two communities can be constructed recursively [35]. This definition enables the specification of the first candidate predictive feature nominated by our predictability assessment: early dispersion of a complex contagion process across network communities should be a reliable predictor that the ultimate reach of the contagion will be significant (see Figure 6).

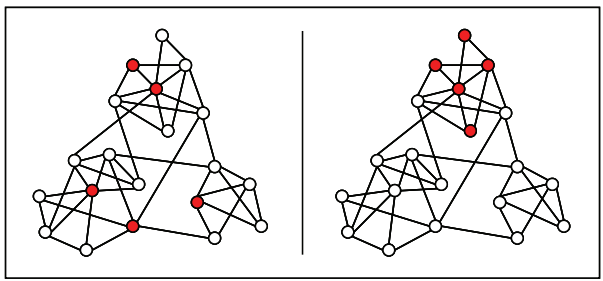

**Figure 6.** Early dispersion across communities is predictive. The cartoon illustrates the predictive feature associated with community structure: contagions initiated with five "seed" individuals are much more likely to propagate widely if the seeds are dispersed across multiple communities (left) rather than concentrated within a single community (right).

We characterize network core-periphery structure in terms of the k-shell decomposition [36]. To partition a network into its k-shells, one first removes all vertices with degree one, repeating this step if necessary until all remaining vertices have degree two or higher; the removed vertices constitute the 1-shell. Continuing in the same way, all vertices with degree two (or less) are recursively removed, creating the 2-shell. This process is repeated until all vertices have been assigned to a k-shell, and the shell with the highest index, the $k_{max}$-shell, is deemed to be the core of the network. This definition permits us to state the second candidate predictive feature nominated via theoretical predictability assessment: early contagion activity within the network $k_{max}$-shell should be a reliable predictor that the reach of the diffusion will be significant (see Figure 7).

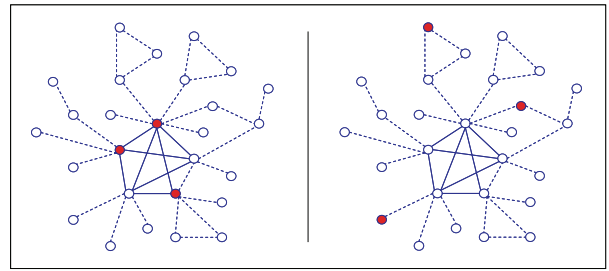

**Figure 7.** Early diffusion within the core is predictive. The cartoon illustrates the predictive feature associated with k-shell structure: contagions initiated with three "seed" individuals are much more likely to propagate widely if the seeds reside within the network's core (left) rather than at its periphery (right).

### C. Prediction Algorithm

Consider the problem of predicting, very early in the lifecycle of a complex contagion event, whether or not the contagion will propagate widely. We adopt an ML approach to this early warning task: given a triggering incident, one or more information sources which reflect the reaction to this trigger by a population of interest, and a specification for what constitutes an "alarming" reaction, the goal is to learn a classifier that accurately predicts, as early as possible, whether or not reaction to the event will eventually become alarming. The ML classifier used in this investigation is the Avatar ensembles of decision trees (A-EDT) algorithm [37]; qualitatively similar results were obtained in tests with other, less sophisticated classifiers [3].

A key step in early warning analysis is determining which characteristics of the phenomenon of interest, if any, possess exploitable predictive power. Based on the results of the preceding predictability assessment study, we consider three general classes of features: 1.) *intrinsics-based features* – measures of the inherent properties and attributes of the "object" being diffused, 2.) *simple dynamics-based features* – metrics which capturing simple properties of the diffusion dynamics (e.g., the rate at which the diffusion is propagating), 3.) *network dynamics-based features* – measures that characterize the way the early diffusion is progressing relative to the network's community and core-periphery structures. Precise definitions for the features in these classes are, of course, application dependent.

The proposed approach to early warning analysis is to identify and collect features from these classes for the event of interest, input the feature values to the A-EDT classifier, and then run the classifier to generate a warning prediction (i.e., a forecast that the event is expected to become 'alarming' or remain 'not alarming'). The algorithm presented below specifies this procedure in general terms, and illustrative instantiations of the procedure are given in the discussions of the case studies. In what follows it is assumed that social media data form the primary source of information concerning the event of interest [38]. However, the analytic process is similar when alternative sources of data are employed [13].

Consider the following early warning algorithm:

**Algorithm EW**

Given: a triggering incident, a definition for what constitutes an 'alarming' reaction, and a set of social media sites (e.g., blogs) B which are relevant to the early warning task.

Initialization: train the A-EDT classifier on a set of events that are qualitatively similar to the triggering event of interest and are labeled as 'alarming' or 'not alarming'.

Procedure:

1. Assemble a lexicon of keywords L that pertain to the triggering event under study.
2. Conduct a sequence of Web crawls and construct a time series of blog graphs $G_B(t)$. For each time period t, label each blog in $G_B(t)$ as 'active' if it contains a post mentioning any of the keyword in L and 'inactive' otherwise.
3. Form the union $G_B = \cup_t G_B(t)$, partition $G_B$ into network communities and into k-shells, and map the partition element structure of $G_B$ back to each of the graphs $G_B(t)$.
4. For each graph $G_B(t)$, compute the values for all features (intrinsics, simple dynamics, and network dynamics).
5. Apply the A-EDT classifier to the time series of features, i.e., the features obtained for the sequence of blog graphs $\{G_B(t_0), …, G_B(t_p)\}$, where $t_0$ and $t_p$ are the triggering event time and present time, respectively. Issue a warning alert if the classifier output is 'alarming'.

We now offer a few remarks concerning Algorithm EW. The keywords in Step 1 can be identified with the help of subject matter experts and also through computational means (e.g., via meme analysis [28]). Step 2 is by now standard, and a variety of tools exist which can perform these tasks [38]. In Step 3, the blog network can be partitioned into communities and k-shells using modularity-based community extraction [35] and standard k-shell decomposition [36], respectively. The particular choices of metrics for the intrinsics, simple dynamics, and network dynamics features computed in Step 4 tend to be problem specific, and typical examples are given in the case studies below. Finally, in Step 5 the feature values obtained in Step 4 serve as inputs to the A-EDT classifier, and the output of the classifier is used to decide whether an alert should be issued.

*D. Meme Case Study*

The goal of this case study is to apply Algorithm EW to the task of predicting whether or not a given *meme* (i.e., short textual phrase which propagates relatively unchanged online) will "go viral". Although it may seem that meme diffusion is not sufficiently costly or controversial to qualify as a complex contagion, [27] shows that *political* memes appear to propagate in this way. Our main source of data on meme dynamics is the dataset archived at the site http://memetracker.org [39] by the authors of [28]. Briefly, the archive [39] contains time series data characterizing the online diffusion of ~70,000 memes during the period between 1 August and 31 December 2008. We are interested in using Algorithm EW to distinguish successful and unsuccessful political memes early in their lifecycle. More precisely, the prediction task is to classify memes into two groups – those which will ultimately be successful (acquire more than S posts) and those that will be unsuccessful (attract fewer than U posts) – very early in the meme lifecycle.

To support an empirical evaluation of the utility of Algorithm EW for this problem, we downloaded from [39] the time series data for slightly more than 70,000 memes. These data contain, for each meme M, a sequence of pairs $(t_1, URL_1)_M, (t_2, URL_2)_M, …, (t_T, URL_T)_M$, where $t_k$ is the time of appearance of the kth blog post or news article that contains at least one mention of meme M, $URL_k$ is the URL of the blog or news site on which that post/article was published, and T is the total number of posts that mention meme M. From this set of time series we randomly selected 100 "successful" political meme trajectories, defined as those corresponding to memes which attracted at least 1000 posts during their lifetimes, and 100 "unsuccessful" political meme trajectories, defined as those whose memes acquired no more than 100 total posts.

Two other forms of data were collected for this study: 1.) a large Web graph which includes websites (URLs) that appear in the meme time series, and 2.) samples of the text surrounding the memes in the posts which contain them. More specifically, we sampled the URLs appearing in the time series for our set of 200 successful and unsuccessful memes and performed a Web crawl that employed these URLs as "seeds". This procedure generated a Web graph, denoted $G_B$, that consists of approximately 550,000 vertices (websites) and 1.4 million edges (hyperlinks), and includes essentially all of the websites which appear in the meme time series. To obtain samples of text surrounding memes in posts, we randomly selected ten posts for each meme and then extracted from each post the paragraph which contains the first mention of the meme.

Algorithm EW employs three types of features: intrinsics, simple dynamics-based, and network dynamics-based. We now describe the instantiation of each of these feature classes for the meme problem. Consider first the intrinsics features, which for the meme application become language-based measures. Each "document" of text surrounding a meme in its (sample) posts is represented by a simple "bag of words" feature vector $x \in \Re^{|V|}$, where the entries of x are the frequencies with which the words in the vocabulary V appear in the document. A language-based feature which might reasonably be expected to be predictive of meme propagation is the sentiment or emotion of documents containing the meme. A simple way to quantify a document's sentiment/emotion is through the use of appropriate lexicons. Let $s \in \Re^{|V|}$ denote a lexicon vector, in which each entry of s is a numerical "score" quantifying the sentiment/emotion intensity of the corresponding word in vocabulary V. The aggregate sentiment/emotion score of document x can then be computed as $score(x) = s^T x / s^T 1$, where 1 is a vector of ones. Thus score(.) estimates document sentiment or emotion as a weighted aver-age of the sentiment or emotion scores for the words comprising the document. (Note that if no sentiment or emotion information is available for a particular word in V then the corresponding entry of s is set to zero.)

To characterize the emotion content of a document we use the Affective Norms for English Words (ANEW) lexicon [40], while positive or negative sentiment is quantified via the "IBM lexicon" [41]. This approach generates four language features for each meme: the happiness, arousal, dominance, and positive/negative sentiment of the sample text surrounding that meme. As a preliminary test, we computed the mean emotion and sentiment of text surrounding the 100 successful and 100 unsuccessful memes in our dataset. On average the text surrounding successful memes is happier, more active, more dominant, and more positive than that surrounding unsuccessful memes (p<0.0001), so it is at least plausible that the language features may possess some predictive power.

Consider next two simple dynamics-based features, defined to capture basic characteristics of the early evolution of meme post volume: 1.) #posts($\tau$) – the cumulative number of posts mentioning the given meme by time $\tau$ (where $\tau$ is small relative to the typical meme lifespan), and 2.) post rate($\tau$) – a simple estimate of the rate of accumulation of these posts at time $\tau$. Recall that predictability assessment suggests that both early dispersion of contagion activity across network communities and early contagion activity within the network core ought to be predictive of meme success. These insights motivate the definition of two network dynamics-based features for meme prediction: 1.) community dispersion($\tau$) – the cumulative number of network communities in the blog graph $G_B$ that, by time $\tau$, contain at least one post which mentions the meme, and 2.) #k-core blogs($\tau$) – the cumulative number of blogs in the $k_{max}$-shell of blog graph $G_B$ that, by time $\tau$, contain at least one post which mentions the meme.

This case study compares the meme early warning accuracy of Algorithm EW, as applied to meme prediction, with that of two other prediction methods: a language-based (LB) strategy and a standard-dynamics (SD) scheme. The LB predictor uses the four language features noted above with the A-EDT classifier to try to distinguish successful and unsuccessful memes, and achieves a prediction accuracy of 66.5% (ten-fold cross-validation). Since simply guessing 'successful' for all memes gives an accuracy of 50%, it can be seen that the language intrinsics, when used alone, possess relatively limited predictive power.

Next we compare the predictive performance of the SD classifier with that of Algorithm EW. The SD predictor combines the four language features with the two simple dynamics features, #posts($\tau$) and post rate($\tau$), within the A-EDT classifier. Because this is representative of state-of-the-art prediction schemes [e.g., 5-8], this approach is referred to as the gold-standard algorithm. The application of Algorithm EW to meme prediction combines the language features with four network dynamics measures: #posts($\tau$), post rate($\tau$), community dispersion($\tau$), and #k-core blogs($\tau$). Sample results from this empirical test are depicted in Figure 8. Each data point represents the average accuracy over ten trials (ten-fold cross-validation). It can be seen from Figure 8 that Algorithm EW outperforms the gold-standard method, especially in the important situation in which it is desired to form predictions soon after the meme is detected. Indeed, these results show that useful predictions can be obtained with Algorithm EW *within the first twelve hours* after a meme is detected (this corresponds to 0.5% of the average meme lifespan). Interestingly, analysis of feature predictive power [3] shows that the most predictive features are, in decreasing order, 1.) community dispersion, 2.) #k-core blogs, 3.) #posts, and 4.) post rate, which supports the conclusions of the complex contagion-based predictability assessment.

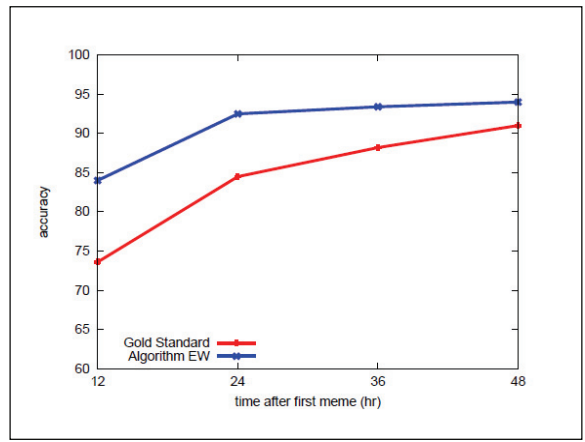

**Figure 8.** Results for meme early warning case study. The plot shows how prediction accuracy (vertical axis) varies with the length of time that has elapsed between meme detection and meme prediction (horizontal axis) for the two classifiers: gold-standard (red) and Algorithm EW (blue).

*E. Early Warning Case Study*

This case study explores the ability of Algorithm EW to provide reliable early warning for politically-motivated distributed denial-of-service (DDoS) attacks, an important class of cyber threats. In particular, we are interested in exploring the utility of Algorithm EW when using social media as an information source. Toward this end, we first identified a set of Internet disruptions which included examples from three distinct classes of activity: 1.) successful DDoS attacks (the events for which we seek early warning; 2.) natural events which disrupt Internet service (disturbances, like earthquakes, that impact the Internet but for which it is known that no early warning signal exists in social media); 3.) quiet periods (periods during which there is social media "chatter" concerning impending DDoS attacks but no successful attacks occurred). Including events selected from these three classes is intended to provide a fairly comprehensive test, as these classes correspond to 1.) the domain of interest, 2.) a set of disruptions which impact the Internet but have no social media warning signal, and 3.) a set of "non-events" which do not impact the Internet but do possess putative social media warning signals.

We selected twenty events from these three classes:

Politically-motivated DDoS attacks:

- Estonia event in April 2007;
- CNN/China incident in April 2008;
- Israel/Palestine conflict event in January 2009;
- DDoS associated with Iranian elections in June 2009;
- WikiLeaks event in November 2010;
- Anonymous v. PayPal, etc. attack in December 2010;
- Anonymous v. HBGary attack in February 2011.

Natural disturbances:
- European power outage in November 2006;
- Taiwan earthquake in December 2006;
- Hurricane Ike in September 2008;
- Mediterranean cable cut in January 2009;
- Taiwan earthquake in March 2010;
- Japan earthquake in March 2011.

Quiet periods:

Seven periods, from 2005 through 2011, during which there were discussions in social media of DDoS attacks on various U.S. government agencies but no successful attacks occurred.

We collected two forms of data for each of these twenty events: *cyber data* and *social data*. The cyber data consist of time series of routing updates which were issued by Internet routers during a one month period surrounding each event. More precisely, these data are the Border Gateway Protocol (BGP) routing updates exchanged between gateway hosts in the Autonomous System network of the Internet. The data were downloaded from the publicly-accessible RIPE collection site [42] using the process described in [43] (see [43] for additional details and background information on BGP routing dynamics). The temporal evolution of the volume of BGP routing updates (e.g., withdrawal messages) gives a coarse-grained measure of the timing and magnitude of large Internet disruptions and thus offers a simple and objective way to characterize the impact of each of the events in our collection. The social data consist of time series of social media mentions of cyber attack-related keywords and memes detected during a one month period surrounding each of the twenty events. These data were gathered using the procedure specified in Algorithm EW.

We apply Algorithm EW to the task of distinguishing the seven DDoS attacks from the thirteen other events in the event set. For simplicity, in this case study we do not use any intrinsics-based features (e.g., language metrics) in the A-EDT classifier, and instead rely upon the four dynamics-based features defined in the meme study. Because the event set in this study includes only twenty incidents, we apply Algorithm EW with two-fold cross-validation. In the case of DDoS events, the blog data made available to Algorithm EW is limited to posts made during the five week period which ended one week before the attack. For the six natural disturbances, the blog data includes all posts collected during the six week period immediately prior to the event, while in the case of the seven non-events, the blog data includes the posts gathered during a six week interval which spans discussions of DDoS attacks on U.S. government agencies.

In this evaluation, Algorithm EW achieves *perfect* accuracy, correctly identifying all 'attack' and 'non-attack' events. If the test is made more difficult, so that the blog data made available to Algorithm EW for attack events is limited to a four week period that ends two weeks before the attack, the proposed approach still achieves 95% accuracy, An examination of the predictive power of the four features used as inputs to the A-EDT classifier reveals that community dispersion is the most predictive measure.

IV. PREDICTIVE DEFENSE

*A. Problem Formulation*

Coevolving adversarial dynamics are present in many application areas, including national security, public policy, business, and economics. There is significant interest to develop *proactive* approaches to dealing with adversarial behavior, in which opponents' future strategies are anticipated and these insights are incorporated into defense designs. Recent work suggests that previous hostile actions and defender responses provide predictive information about future attacker behavior because of the coevolving relationship that exists between adversaries [44,21]. However, little is known about how to systematically exploit this predictability to design effective countermeasures. This section considers the following concrete instantiation of the proactive defense problem: given some (possibly limited) history of attacker actions, design a predictive defense system which performs well against both current and future attacks.

It is reasonable to expect that concepts and techniques from *game theory* [29,30] might be useful in designing predictive defense systems, and indeed such approaches have been attempted in different domains (see [21,45] for relevant background). These investigations, although useful, have not tended to produce practically-implementable results. While there are several challenges to successfully applying game-theoretic methods to adversarial dynamics of real-world scale and complexity, we mention two that have been particularly daunting. First, the space of possible attacker actions is typically very large in realistic environments. Because the complexity of game models usually increases exponentially with the number of actions available to the players [29,45], this has made game models intractable in practice. And second, it has proved difficult to develop models that capture evolving attacker behavior in any but the most idealized (and unrealistic) situations.

In this section we overcome these challenges by approaching the task of developing game-based models for adversarial interactions from a sociological perspective [30]. More specifically, we develop a simple game model which explicitly leverages empirical data directly within an ML framework, enabling adversary behavior to be predicted and countered in realistic settings. Crucially, the proposed approach derives the optimal defense for the predicted attacks, rather than attempting to obtain perfect predictions, and therefore enjoys robust performance in the presence of (inevitable) prediction errors. (See, for instance, [30] for an argument for the need to explicitly connect game models with data if game theory is to impact the empirically-oriented science of sociology, and for suggestions concerning some ways sociology might contribute to an empirical game theory.)

We approach the task of countering adversarial behavior as an ML classification problem, in which the objective is to distinguish innocent and malicious activity. Each instance of activity is represented as a feature vector $x \in \Re^{|F|}$, where entry $x_i$ of x is the value of feature i for this instance and F is the set of instance features. In what follows, F is a set of "reduced" features, obtained by projecting the original feature vectors into a lower-dimensional space. While feature reduction is standard practice in ML [3], we show below that *aggressive* reduction allows us to efficiently manage the complexity of our game models. Behavior instances x belong to one of two classes: positive/malicious and negative/innocent (generalizing to more than two behavior classes is straightforward [3]). The goal is to learn a vector $w \in \Re^{|F|}$ such that classifier orient = sign($w^T x$) accurately estimates the class of behavior x, returning +1 (−1) for malicious (innocent) activity.

*B. Predictability Assessment*

As indicated in Section III, it is useful to assess the predictability of a phenomenon before attempting to predict its evolution, for example to identify measurables which possess predictive power [1,13]. There has been limited theoretical work assessing predictability of adversarial dynamics, but existing studies suggest attack-defend coevolution often generates predictable dynamics. For instance, although [46] finds that, in some repeated games, certain player strategies lead to chaotic dynamics, [22] shows a large range of player strategies in repeated two-player and multi-player games result in predictable adversarial behavior. Here we supplement this theoretical work by conducting an empirical investigation of predictability, and select as our case study a cyber security problem – Spam filtering – which possesses attributes that are representative of many adversarial domains.

To conduct this investigation, we first obtained a large collection of emails from various publicly-available sources for the period 1999-2006, and added to this corpus a set of Spam emails acquired from B. Guenter's Spam trap for the same time period. Following standard practice, each email is modeled as a "bag of words" feature vector $x \in \Re^{|F|}$, where the entries of x are the frequencies with which the words in vocabulary F appear in the message. The resulting dataset consists of ~128,000 emails composed of more than 250,000 features. We extracted from this collection of Spam and non-Spam emails the set of messages sent during the 30 month period between January 2001 and July 2003 (other periods exhibit very similar behavior). Finally, the dimension of the email feature space is reduced via singular value decomposition (SVD) analysis [3], yielding a reduction in feature space dimension |F| of four orders of magnitude (from ~250K to 20).

We wish to examine, in an intuitive way, the predictability of Spammer adaptation, and propose two simple but reasonable criteria with which to empirically evaluate predictability: *sensibility* and *regularity* (a more systematic and mathematically-rigorous frameworks for defining and assessing predictability is derived in [1,13] and summarized in Section III). More specifically, and in the context of Spam, it would be *sensible* for Spammers to adapt their messages over time in such a way that Spam feature vectors $x_S$ become more like feature vectors $x_{NS}$ of legitimate emails, and *regularity* in this adaptation might imply that the values of the individual elements of $x_S$ approach those of $x_{NS}$ monotonically.

To permit convenient examination of the evolution of feature vectors $x_S$ and $x_{NS}$ during the 30 month period under study, the emails are first binned by quarter. Next, the average values for each of the 20 (reduced dimension) features are computed for all Spam emails and all non-Spam emails (separately) for each quarter. Figure 9 illustrates the feature space dynamics of Spam and non-Spam messages for one representative element (F1) of this reduced feature space. As seen in the plot, the value of feature F1 for Spam approaches the value of this feature for non-Spam, and this increasing similarity is a consequence of changes in the composition of Spam messages (the value of F1 for non-Spam emails is essentially constant). The dynamics of the other feature values (not shown) are analogous.

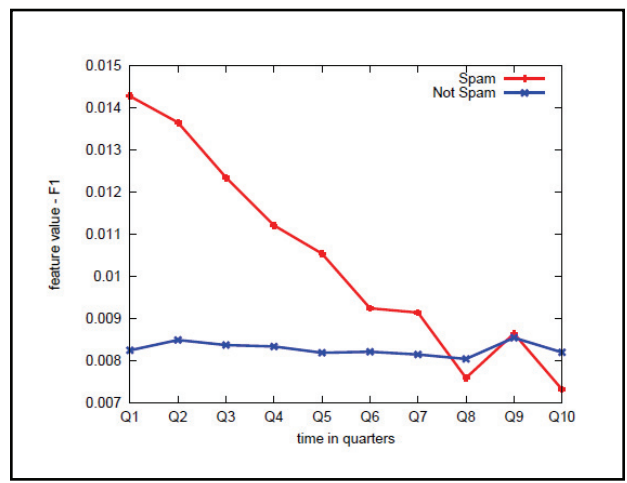

**Figure 9.** Spam/non-Spam evolution in feature space. The plot depicts evolution of feature F1 for Spam (red) and non-Spam (blue) during the ten quarters of the study.

Observe that the Spam dynamics illustrated in Figure 9 reflect *sensible* adaptation on the part of Spammers: the features of Spam email messages evolve to appear more like those of non-Spam email and therefore to be more difficult to identify. Additionally, this evolution is *regular*, with feature values for Spam approaching those for non-Spam in a nearly-monotonic fashion. Thus this empirical analysis indicates that coevolving Spammer-Spam filter dynamics possesses some degree of predictability, and that the features employed in Spam analysis may have predictive power; this result is in agreement with the conclusions of the theoretical predictability analysis reported in [22]. Moreover, because many of the characteristics of Spam-Spam defense coevolution are also present in other adversarial systems, this result suggests these other systems may have exploitable levels of predictability as well.

*C. Predictive Defense*

The proposed approach to designing a predictive defense system which works well against both current and future attacks is to combine ML with a simple game-based model for adversary behavior. In order to apply game-theoretic methods, it is necessary to overcome the complexity and model-realism challenges mentioned above. We address problem complexity by modeling adversary actions directly in an aggressively-reduced feature space, so that the (effective) space of possible adversary actions which must be considered is dramatically decreased. The difficulty of deriving realistic representations for attacker behavior is overcome by recognizing that the actions of attackers can be modeled as attempts to *transform* data (i.e., feature vectors x) in such a way that malicious and innocent activities are indistinguishable. (This is in contrast to trying to model the attack instances "from scratch"). It is possible to model attacker actions as transformations of data because, within an empirical ML problem formulation, historical attack data are available in the form of training instances.

We model adversarial coevolution as a sequential game, in which attacker and defender iteratively optimize the following objective function:

$$\min_{w} \max_{a} \left[ -\alpha \|a\|^3 + \beta \|w\|^3 + \sum_{i} \text{loss}(y_i, w^T(x_i + a)) \right] \quad (5)$$

In (5), the loss function represents the misclassification rate for the defense system, where $\{y_i, x_i\}_{i=1}^{n}$ denotes pairs of activity instances $x_i$ and labels $y_i$, and vector w parameterizes the defense (recall that the defense attempts to distinguish malicious and innocent activity using the classifier orient = sign($w^T x$)). The attacker attempts to circumvent the defense by transforming the data through vector $a \in \mathfrak{R}^{|F|}$, and the defender's goal is to counter this attack by appropriately specifying classifier vector $w \in \mathfrak{R}^{|F|}$. The terms $-\alpha \|a\|^3$ and $\beta \|w\|^3$ define "regularizations" imposed on attacker and defender actions, respectively, as discussed below.

Observe that (5) models the attacker as acting to increase the misclassification rate with vector a, subject to the need to limit the magnitude of this vector (large a is penalized via the term $-\alpha \|a\|^3$). This model thus captures in a simple way the fact that the actions of the attacker are in reality always constrained by the goals of the attack. For instance, in the case of Spam email attacks, the Spammer tries to manipulate message x in such a way that it "looks like" legitimate email and evades the Spam filter w. However, transformed message x+a must still communicate the desired information to the recipient or the attacker's goal will not be realized, and so the transformation vector a cannot be chosen arbitrarily.

The defender attempts to reduce the misclassification rate with an optimal choice for vector w, and avoids "over-fitting" through regularization with the $\beta \|w\|^3$ term [3]. Notice that the formulation (5) permits the attacker's goal to be modeled as counter to, but not exactly the opposite of, the defender's goal, and this is consistent with many real-world settings. Returning to the Spam example, the Spammer's objective of delivering messages which induce profitable user responses is not the inverse of an email service provider's goal of achieving high Spam recognition with a very low false-positive rate.

The preceding development can be summarized by stating the following predictive defense (PD) algorithm:

**Algorithm PD**

1. Collect historical data $\{y_i, x_i\}_{i=1}^{n}$ which reflects past behavior of the attacker and past legitimate behavior.
2. Optimize objective function (5) to obtain the predicted actions $a \in \mathfrak{R}^{|F|}$ of the attacker and the optimal defense $w \in \mathfrak{R}^{|F|}$ to counter this attack.
3. Estimate the status of any new activity x as either malicious (+1) or innocent (−1) via orient = sign($w^T x$).

Observe that Step 2 of this algorithm can be interpreted as first predicting the attacker strategy through computation of attack vector a, and then learning an appropriate countermeasure w by applying ML to the "transformed" data $\{y_i, x_i + a\}_{i=1}^{n}$.

*D. Spam Case Study*

This case study examines the performance of Algorithm PD for the Spam filtering problem. We use the Spam/non-Spam email dataset introduced above, consisting of ~128,000 messages that were sent during the period 1999-2006. The study compares the effectiveness of Algorithm PD, implemented as a Spam filter, with that of a well-tuned naïve Bayes (NB) Spam filter [21]. Because NB filters are widely used and work very well in Spam applications, this filter is referred to as the gold-standard algorithm. We extract from our dataset the 1000 oldest legitimate emails and 1000 oldest Spam messages for use in training both Algorithm PD and the gold-standard algorithm. The email messages sent during the four year period immediately following the date of the last training email are used as test data. More specifically, these emails are binned by quarter and then randomly sub-sampled to create balanced datasets of Spam and legitimate emails for each of the 16 quarters in the test period.

Recall that Algorithm PD employs aggressive feature space dimension reduction to manage the complexity of the game-based modeling process. This dimension reduction is accomplished here through SVD analysis, which reduces the dimension |F| of feature vectors from ~250K to 20) [3]. (The orthogonal basis used for this reduction is derived by performing SVD analysis using the 1000 non-Spam and 1000 Spam training data.) We remark that good classification accuracy can be obtained with a wide range of (reduced) feature space dimensions. For example, a Spam filter accuracy of ~97% is achieved with the training data when using an NB classifier implemented with a feature space dimension ranging from |F|=100,000 to |F|=5 (accuracy is estimated via two-fold cross-validation).

The gold-standard strategy is applied as described in [21]. Algorithm PD is implemented with parameter values $\alpha = 0.001$ and $\beta = 0.1$, and with a sum-of-squares loss function. As discussed above, motivations for developing predictive defenses include the capability to perform well against new attacks and the possibility to maintain an effective defense without the need for frequent retraining, which is ordinarily expensive and time-consuming. To explore these issues, in this case study we train Algorithm PD and the gold-standard algorithm *once,* using the 1000 legitimate/1000 Spam dataset, and then apply the filters without retraining on the first four years of emails that follow these 2000 emails.

Sample results from this study are depicted in Figure 10. Each data point in the plots represents the average accuracy over ten trials (two-fold cross-validation). It can be seen that the filter based upon Algorithm PD significantly outperforms the gold-standard method: the predictive defense experiences almost no degradation in filtering accuracy over the four years of the study, while the gold-standard method suffers a substantial drop in accuracy during this period. These results suggest that the combination of ML and game theory-based adversary model offers an effective means of defending against new attacks. Additionally, the results indicate that predictive defense permits good performance to be realized with much less retraining than is usually required.

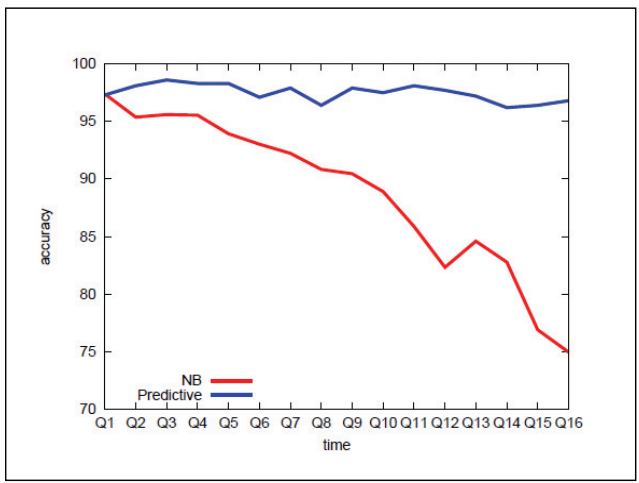

**Figure 10.** Results for Spam filtering case study. The plot shows how filter accuracy (vertical axis) varies with time (horizontal axis) for the gold-standard classifier (red) and Algorithm PD (blue).

### E. Randomized Feature Learning

An important consideration when applying ML techniques in adversarial settings is the extent to which adversaries can "reverse-engineer" the learning algorithm and use these insights to circumvent the classifier. One way to increase the difficulty of the adversary's reverse engineering task is to employ "randomized feature" learning [47]. Here we explore in a preliminary way the following three-step implementation of this idea: 1.) divide the set of available features into randomly-selected, possibly overlapping subsets; 2.) train one classifier for each subset of features; and 3.) alternate between classifiers in a random fashion during operation. The fact that good classifier performance is often obtainable with only a few features (see the Spam example above) suggests the feasibility of employing multiple small subsets of randomly-selected features in a classifier.

To test the effectiveness of this strategy, we use a variant of the optimization process specified in (1). More specifically, we first use training data $\{y_i, x_i\}_{i=1}^n$ to computed the classifier vector w in two ways: 1.) using the full set of (reduced-dimension) features F, 2.) using two subsets of features randomly selected from set F; the resulting classifier vectors are denoted $w_F$ and $\{w_{F1}, w_{F2}\}$. (1) is then employed to compute the optimal attack against classifier vector $w_F$, denoted $a_F$, and to compute the optimal attack for the defense consisting of randomly alternating classifiers $w_{F1}$ and $w_{F2}$, designated $a_{F12}$.

Applying this evaluation process to the 2000 email training dataset described in Section IVD suggests randomized feature leaning may be an effective way to reduce the efficacy of adversary reverse engineering methods. We define F to be the set of 20 features with largest singular values, and build sets F1 and F2 by randomly sampling F (with replacement) until each subset contains 10 features. The classification accuracy of $w_F$ against *nominal* data (i.e., with a=0) is superior to that provided by a classifier which randomly alternates between classifiers $w_{F1}$ and $w_{F2}$, but the difference is modest – the accuracies are 98.4% and 96.2%, respectively, for $w_F$ and $\{w_{F1},w_{F2}\}$ (two-fold cross-validation). Crucially, however, the randomized feature classifier is substantially more robust against *attack* data (i.e., data for $a=a_F$ or $a=a_{F12}$). Indeed, the accuracy of classifier $w_F$ is only 66.1% against attack data, while the accuracy of randomized feature classifier $\{w_{F1},w_{F2}\}$ drops less than 10%, to 86.8%, in this setting (two-fold cross-validation). Figure 11 summarizes these results.

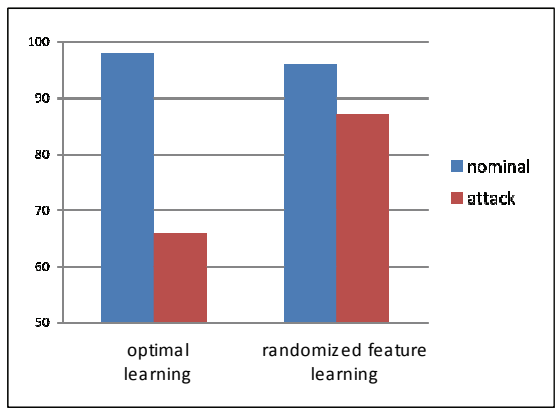

**Figure 11.** Results for randomized feature learning case study. Bar charts show Spam-non-Spam classification accuracy for classifiers $w_F$ (left bars) and $\{w_{F1}, w_{F2}\}$ (right bars) for nominal Spam (blue) and "attack" Spam (red) (two-fold cross-validation).

### V. SUMMARY

This paper proposes that predictive analysis techniques can be improved by leveraging sociological models, and explores this possibility by considering three challenging prediction tasks: 1.) inferring the signs (friendly or antagonistic) of ties in social networks, 2.) predicting whether an emerging social diffusion event will propagate widely or quickly dissipate, and 3.) anticipating and defending future actions of opponents in adversarial settings. In each case, we derive a novel machine learning-based prediction algorithm which incorporates a sociological model in its development and show that the new algorithm outperforms a "gold-standard" method in empirical tests. Tak-en together, the collection of results presented in this paper indicate that incorporating simple models from sociology can substantially improve the performance of prediction methods, particularly in applications in which there is only limited data available for training and implementing the algorithms.

ACKNOWLEDGEMENTS

This work was supported by the U.S. Department of Defense, The Boeing Company, and the Laboratory Directed Research and Development Program at Sandia National Laboratories. We thank Chip Willard of the U.S. Department of Defense, Curtis Johnson of Sandia National Laboratories, and Anne Kao of Boeing for numerous helpful discussions on aspects of this research.

REFERENCES

[1] Colbaugh, R. and K. Glass, "Predictive analysis for social processes I: Multi-scale hybrid system modeling, and II: Predictability and warning analysis", *Proc. 2009 IEEE Multi-Conference on Systems and Control*, Saint Petersburg, Russia, July 2009.

[2] Choi, H. and H. Varian, "Predicting the present with Google Trends", SSRN Preprint, April 2009.

[3] Hastie, T., R. Tibshirani, and J. Friedman, *The Elements of Statistical Learning*, Second Edition, Springer, New York, 2009.

[4] Colbaugh, R., K. Glass, and P. Ormerod, "Predictability of 'unpredictable' cultural markets", *105th Annual Meeting of the American Sociological Association*, Atlanta, GA, August 2010.

[5] Asur, S. and B. Huberman, "Predicting the future with social media", *Proc. IEEE/WIC/ACM International Conference on Web Intelligence and Intelligent Agent Technology*, Toronto, Ontario, Canada, September 2010.

[6] Goel, S., J. Hofman, S. Lahaie, D. Pennock, and D. Watts, "Predicting consumer behavior with Web search", *Proc. National Academy of Sciences USA*, Vol. 107, pp. 17486-17490, 2010.

[7] Lerman, K. and T. Hogg, "Using stochastic models to describe and predict social dynamics of Web users", arXiv preprint, October 2010.

[8] Bollen, J., H. Mao, and X. Zeng, "Twitter mood predicts the stock market", *J. Computational Science*, Vol. 2, pp. 1-8, 2011.

[9] Colbaugh, R. and K. Glass, "Detecting emerging topics and trends via predictive analysis of 'meme' dynamics", *Proc. 2011 AAAI Spring Symposium Series*, Palo Alto, CA, March 2011.

[10] Lui, C., P. Metaxas, and E. Mustafaraj, "On the predictability of the U.S. elections through search volume activity", *Proc. IADIS e-Society Conference*, Avila, Spain, March 2011.

[11] Guman, G., "Internet search behavior as economic forecasting tool: The case of inflation expectations", *J. Economic and Social Measurement*, Vol. 36, pp. 119-167, 2011.

[12] Amodea, G., R. Blanco, and U. Brefeld, "Hybrid models for future event prediction", *Proc. CIKM '11*, Glasgow, Scotland, October 2011.

[13] Colbaugh, R. and K. Glass, "Early warning analysis for social diffusion events", *Security Informatics*, Vol. 2, 2012, in press.

[14] Clauset, A., C. Moore, and M. Newman, "Hierarchical structure and the prediction of missing links in networks", *Nature*, Vol. 453, pp. 98-101, 2008.

[15] Pang, B. and L. Lee, "Opinion mining and sentiment analysis", *Foundations and Trends in Information Retrieval*, Vol. 2 , pp. 1-135, 2008.

[16] Abbasi, A., H. Chen, and A. Salem, "Sentiment analysis in multiple languages: Feature selection for opinion classification in Web forums", *ACM Transactions on Information Systems*, Vol. 26, pp. 1-34, 2008.

[17] Lampos, V., T. De Bie, and N. Cristianini, "Flu detector – Tracking epidemics on Twitter, *ECML PKDD 2010*, Springer LNAI 6323, 2010.

[18] Christakis, N. and J. Fowler, "Social network sensors for early detection of contagious outbreaks", *PLoS ONE*, Vol. 5, e12948, 2010.

[19] Ayers, J., K. Ribisi, and J. Brownstein, "Tracking the rise in popularity of electronic nicotine delivery systems using search query surveillance", *American J. Preventative Medicine*, Vol. 41, pp. 1-6, 2011.

[20] Glass, K. and R. Colbaugh, "Estimating the sentiment of social media content for security informatics applications", *Security Informatics*, Vol. 1, No. 3, pp. 1-16, 2012.

[21] Colbaugh, R. and K. Glass, "Proactive defense for evolving cyber threats", *Proc. 2011 IEEE International Conference on Intelligence and Security Informatics*, Beijing, China, July 2011.

[22] Colbaugh, R., "Arctic ice, George Clooney, lipstick on a pig, and insomniac fruit flies: Combining kd and m&s for predictive analysis", *Proc. ACM KDD '11*, San Diego, CA, August 2011.

[23] Heider, F., "Attitude and cognitive organization", *J. Psychology*, Vol. 21, pp. 107-112, 1946.

[24] Cartwright, D. and F. Harary, "Structural balance: A generalization of Heider's theory", *Psychological Review*, Vol. 63, pp. 277-293, 1956.

[25] Leskovec, J., D. Huttenlocher, and J. Kleinberg, "Predicting positive and negative links in online social networks", *Proc WWW 2010*, Raleigh, NC, April 2010.

[26] Centola, D., "The spread of behavior in an online social network experiment", *Science*, Vol. 329, pp. 1194-1197, 2010.

[27] Romero, D., B. Meeder, and J. Kleinberg, "Differences in the mechanics of information diffusion across topics: Idioms, political hashtags, and complex contagion on Twitter", *Proc WWW 2011*, Hyderabad, India, March 2011.

[28] Leskovec, J., L. Backstrom, and J. Kleinberg, "Meme-tracking and the dynamics of the news cycle", *Proc. ACM KDD '09*, Paris, France, June 2009.

[29] von Neumann, J. and O. Morgenstern, *Theory of Games and Economic Behavior*, Princeton University Press, 1944.

[30] Swedberg, R., "Sociology and game theory: Contemporary and historic perspectives", *Theory and Society*, Vol. 30, pp. 301-335, 2001.

[31] Maniu, S., B. Cautis, and T. Abdessalem, "Building a signed network from interactions in Wikipedia", *Proc. DBsocial '11*, Athens, Greece, June 2011.

[32] Marvel, S., J. Kleinberg, R. Kleinberg, and S. Strogatz, "Continuous-time model of structural balance", *Proc. National Academy of Sciences USA*, Vol. 108, pp. 1771-1776, 2011.

[33] Axelrod, R. and D. Bennett, "Landscape theory of aggregation", *British J. Political Science*, Vol. 23, pp. 211-233, 1993.

[34] Zachary, W., "Information flow model for conflict and fission", *J. Anthropological Research*, Vol. 33, pp. 452-473, 1977.

[35] Newman, M., "Modularity and community structure in networks", *Proc. National Academy of Sciences USA*, Vol. 103, pp. 8577-8582, 2006.

[36] Carmi, S., S. Havlin, S. Kirkpatrick, Y. Shavitt, and E. Shir, "A model of Internet topology using the k-shell decomposition", *Proc. National Academy of Sciences USA,* Vol. 104, pp. 11150-11154, 2007.

[37] http://www.sandia.gov/avatar/, accessed July 2010.

[38] Glass, K. and R. Colbaugh, "Web analytics for security informatics", *Proc. IEEE European Intelligence and Security Informatics Conference*, Athens, Greece, September 2011.

[39] http://memetracker.org, accessed January 2010.

[40] Bradley, M. and P. Lang, "Affective norms for English words (ANEW): Stimuli, instruction manual, and affective ratings", Technical Report C1, University of Florida, 1999.

[41] Ramakrishnan, G., A. Jadhav, A. Joshi, S. Chakrabarti, and P. Bhattacharyya, "Question answering via Bayesian inference on lexical relations", *Proc. Annual Meeting of the Association for Computational Linguistics*, Sapporo, Japan, July 2003.

[42] http://data.ris.ripe.net/, last accessed July 2011.

[43] Glass, K., R. Colbaugh, and M. Planck, "Automatically identifying the sources of large Internet events", *Proc. IEEE International Conference on Intelligence and Security Informatics*, Vancouver, Canada, May 2010.

[44] Bozorgi, M., L. Saul, S. Savage, and G. Voelker, "Beyond heuristics: Learning to classify vulnerabilities and predict exploits", *Proc. ACM KDD '10*, Washington DC, July 2010.

[45] Colbaugh, R. and K. Glass, "Predictive defense against evolving adversaries", SAND Report, Sandia National Laboratories, December 2011.

[46] Sato, Y., E. Akiyama, and J.D. Farmer, "Chaos in learning a simple two-person game", *Proc. National Academy of Sciences USA,* Vol. 99, pp. 4748-4751, 2002.

[47] Johnson, C., Personal communication, December 2011.